\documentclass[acs,pre,twocolumn,showpacs,floatfix,preprintnumbers,superscriptaddress,amsmath,amssymb]{revtex4-1}
\usepackage{amsmath}
\usepackage{amssymb}
\usepackage{graphicx}
\usepackage{epsfig}
\usepackage{dcolumn}
\usepackage{bm}
\usepackage{array}
\usepackage{mathtools}
\usepackage{epstopdf}
\usepackage{setspace}
\usepackage{color,soul}
\begin{document}
\title{Elasticity of Single Flexible Polymer Chains in Good and Poor Solvents}
\author{Vikhyaat Ahlawat}
\affiliation{Department of Physics, Indian Institute of Science Education and Research, 
Dr. Homi Bhabha Road, Pune 411 008, India}
\author{Shatruhan Singh Rajput}
\author{Shivprasad Patil}
\email{s.patil@iiserpune.ac.in}
\affiliation{Department of Physics, Indian Institute of Science Education and Research, 
Dr. Homi Bhabha Road, Pune 411 008, India}

\begin{abstract}
\section*{Abstract}
Force versus extension curves measure entropic elasticity of single polymer chain in force spectroscopy experiments. A Worm-like Chain model is used to describe force extension experiments with an intrinsic chain parameter called persistence length, which is a measure of local bending flexibility. For flexible polymers, there is a discrepancy in estimates of persistence length in various force regimes. For instance, Atomic Force Microscopy (AFM) based pulling experiments report anomaly low values which are also inconsistent with magnetic tweezers experiments. 
To understand this, we investigate the role of coupling between microscopic force probe and intrinsic elasticity of polyethylene glycol chain in AFM-based experiments. We perform experiments using oscillatory rheology by providing an external excitation of fixed frequency to the probe. We show that a proper quantification of elastic response measured directly by oscillatory technique deviates significantly from conventional force-extension curves. The persistence length obtained by fitting WLC to stiffness extension data  matches well with equilibrium tweezers experiments. 
In addition, for polystyrene chain in poor solvent no deviation in elastic response is observed between oscillatory and constant velocity pulling experiments. However, such deviation is seen for polystyrene in good solvent. We attribute this to hydrophobic interaction between monomers of polystyrene in water. Our results suggest that oscillatory rheology on single polymer chains provide quantitative estimate of its elastic response.  The consistency in values of persistence length using magnetic tweezers experiments in low force regime and the AFM experiments in high force regime suggests that WLC is successful in describing the polymer elasticity in the force range typically probed in AFM experiments.  
\end{abstract}
\maketitle

\section*{Introduction}
Elasticity of single polymer chain under external force is a fundamental problem in polymer science\cite{colby}. It has been studied in the context of understanding mechanical properties of complex biological and polymeric systems\cite{carrillo} to mechanobiology of protein titin in muscle contraction\cite{eckels}. With wide technological importance, it plays crucial role in the field of microrheology\cite{charles} and fundamentals of protein collapse and protein folding\cite{zoldak}. Polymer elasticity is largely entropic in nature and it arises from changes in vast conformational space of polymer's backbone dihedral angles as external force is applied.

Single molecule force spectroscopy techniques such as atomic force microscopy (AFM), optical Tweezers (OTs) and magnetic tweezers (MTs) measures elasticity by generating force versus extension curves with piconewton force resolution and nanometer spatial resolution\cite{neumann}. Previous studies based on single molecule force spectroscopy have been done on synthetic polymers and biopolymers\cite{vancso,Cui} including DNA\cite{Bustamante}, polysaccharides\cite{Marszalek} and proteins\cite{Lorna}. At low forces($<$ 10 pN), chain extension is linear with force(x $\sim$ f) behaving as ideal chain in theta solvent or as a real swollen chain(x $\sim$ f$^{2/3}$) in good solvent. As force is increased, extension changes nonlinearly with force and asymptotically approaches its contour length. This nonlinear stretching regime is described by two main classes of entropic models that include worm-like chain (WLC){\cite{Marko}} and freely jointed chain(FJC)\cite{colby} models. WLC treats the polymer as a continuum structureless chain and persistence length ($l_p$), a measure of local bending stiffness of chain, is estimated as a fitting parameter. In force regime f $>$ $k_BT/l_p$, elasticity of DNA\mbox{\cite{bustamante1,omar1}}, proteins\mbox{\cite{popa,berne}} and synthetic polymer chains \mbox{\cite{saleh,saleh3}} is described by WLC. Typically these experiments were carried out with magnetic tweezers(MTs) in the force range of 1-100 pN. In contrast to MT, AFM experiments done in high force regime of 20-500 pN show anomaly low and unphysical values of persistence length \mbox{\cite{vancso,walker,zhang,zhang2}}. To justify such low values, models such as FJC and its modifications were chosen to describe force-extension curve \cite{radiom,Cui,vancso}. It is expected that beyond a threshold force, the nonlinear stretching regime will transition from continuum WLC to a behavior dominated by discrete nature of polymer chain(FJC)\mbox{\cite{rosa,rubinstein}}. However, such a transition has not been reported in AFM experiments which covers the range of higher forces\cite{saleh3}.

In force spectroscopy experiments, one end of the polymer is typically anchored to a fixed surface with other end tethered to a macroscopic force probe like AFM cantilever or a bead. To generate force-extension curves either the end-to-end distance is controlled externally and
force is measured or vice versa. The choice of external
control or a statistical ensemble reflects the complexity
of interpreting the experimental data. In a typical AFM setup, force is measured indirectly through deflection of the cantilever as end-to-end distance is varied at constant velocity. Such experiments correspond to an ensemble
of coupled cantilever-polymer system, wherein polymer
stiffness and cantilever probe act simultaneously in an
intricate way\mbox{\cite{franco,kreuzer,staple}}. This convolution is likely to produce a biased molecular trajectory in which derivative
of force-extension curve is not the intrinsic stiffness of
the polymer. In MT experiments however, the force is controlled externally and kept fixed while position of the bead is monitored. Such experiments measure properties of an isolated polymer in equilibrium, averaging out the effect of macroscopic probe\cite{kreuzer,staple}.
In literature, method such as equilibrium Weighted Histogram Analysis Method(WHAM)\cite{franco,kumar} and others\cite{thirumalai,netz} have been proposed to deconvolute the apparatus effects and obtain the intrinsic trajectory of the molecule. 

Oscillatory rheology on composite mechanical system is an alternative method which may allow deconvolution\cite{netz}. It is based on the idea of differentiating the dynamical linear response of a single component from the overall response of composite system. In AFM experiments, it can be implemented on cantilever plus polymer system by adding oscillatory perturbation to cantilever probe. In this case, linear response of cantilever and the polymer due to small perturbation is additive to the overall system response. Here, we show that such additivity in linear response allows differentiation of polymer intrinsic response from effects arising due to its coupling to cantilever probe. We implement this by providing oscillations to AFM cantilever while the polymer is simultaneously pulled at constant velocity. Extracting in-phase and out-of-phase linear response to this perturbation, we propose a method to determine the elasticity of single polymer chain. It is demonstrated that elasticity determined from in-phase response show significant deviation from constant velocity pulling experiments. The elastic response of synthetic flexible polymers such as Polyethylene Glycol(PEG) and Polystyrene(PS) is extracted from dynamic oscillatory method and analysed with WLC. For PEG in good solvents, persistence length is large ($\sim$0.64 nm) compared to constant velocity pulling experiments ($\sim$0.13 nm). This value is physically reasonable and matches with equilibrium measurement carried out with MT. 

In addition, the effect of solvent quality on dynamic oscillatory method is considered. The poor solvent can significantly alter the polymer dynamics and elasticity \cite{halperin,radiom,grater,berne2}. For polymers in poor solvents, such as PS in water, the deviation in stiffness measured using constant velocity and oscillatory method is not observed. This is in contrast to good solvent(8M urea) for PS where marked deviation is observed with persistence length about three times ($\sim$0.8 nm) compared to pulling experiments ($\sim$0.25 nm). Although, there are many organic liquids which are typically used as good solvent for PS, we resorted to use  8M urea since it is feasible in AFM experiments compared to volatile organic solvents. Deviation is possibly due to additional hydrophobic forces operating for polystyrene in water. Our results, for the first time, reconcile AFM and  MT measurements of single chain elasticity in high force regime (f $>$ $k_BT/l_p$) and confirms that WLC is largely successful in the force regime probed using AFM. However, the results hint at using WLC with caution for polymers in poor solvents. 

\section*{Experimental Section}

\textbf{Sample preparation for PEG :} Monofunctionalized poly(ethylene) glycol (MeO-PEG-SH) with a molecular weight of 10 kDa was purchased from Sigma-Aldrich with one end functionalized with a thiol group (-SH). For measurements in water Poly(ethylene) glycol was dissolved in deionized-water($>$16 M$\Omega$ cm) to prepare a solution with concentration of 20 $\mu$M. Similar concentration was used to prepare a solution for PEG in 2-propanol. Thermally evaporated gold coverslips were prepared. Gold is inert and does not react with anything chemically but accumulates many organic contaminants due to its high surface free energy. To clean such a gold surface we used UV ozone treatment which is known to remove organic entities and prepare clean gold surfaces. Before use, gold coverslips were gently rinsed with ethanol and deionized water to remove salts. A drop of  60 $\mu$L solution of PEG in respective solvents was deposited on gold cover-slip and incubated for 20 min for strong attachment of thiol ended PEG group on gold surface. Thereafter, the solution was thoroughly rinsed with respective solvents removing any unbound PEG and other impurities. Sample was mounted in the fluid cell for measurement in deionized(DI) water or 2-propanol. 
\par
\textbf{Sample preparation for Polystyrene}
Polystyrene of molecular weight 192 kDa was purchased from Sigma-Aldrich. It was dissolved and stirred in THF$($tetrahydrofuran$)$ for 1 hour with a concentration of $0.1 \mu$M.  A hot piranha solution$($4:1 mixture of concentrated sulphuric acid and hydrogen peroxide$)$ treated glass coverslip was rinsed with ethanol and deionized$($DI$)$ water. Thereafter, 50 $\mu$l of the solution incubated on it for 10 min. To remove loosely adsorbed polymer, sample was rinsed thoroughly with THF and dried for half an hour. Final sample was mounted in fluid cell and filled with deionized(DI)water or 8M urea for measurement.

\subsection*{Force measurements} The force measurements were carried out on a commercial atomic force microscope (Model JPK  Nanowizard II, Berlin, Germany). Gold coated cantilevers from MikroMasch SPM probes were used for experiment. The spring constant of the cantilever was  calibrated using thermal fluctuation method\cite{Butt}. Cantilever was hard pressed against a clean glass coverslip and from already calibrated z- piezo movement sensitivity in units of nm/V was determined. This sensitivity was then used in thermal fluctuation method to calculate spring constant. The spring constant of the cantilever used in the measurement was in the range $0.6-0.8$  N/m with resonance frequency $\sim$ 13 KHz. Such a range was found appropriate for both force and dynamic measurement. Force versus extension profiles were generated following a standard procedure. Initially the cantilever was approached to substrate with a contact setpoint of about 2 nN. Contact setpoint was maintained for 1-2 seconds allowing adsorption of polymer strand on cantilever tip. Cantilever deflects as the attached polymer is stretched while the cantilever was pulled with constant velocity of 150 nm/s. Both z-piezo distance and cantilever deflection detected by quadrant photodiode are recorded. Extension in z-piezo distance is corrected for vertical deflection and force is obtained by multiplying the deflection by spring constant of cantilever. In force spectroscopy experiments, it is crucial to record only single molecule events since the binding between the tip and polymer is non-specific. To achieve this:(1) sample was prepared with low concentration of polymer solution and rinsed several times before measurement. This ensured that mostly individual binding events were captured in force profile. We have discarded any multiple events from analysis and included data which shows cantilever picking only single polymer chain. (2) The  curves were rescaled with their apparent contour length obtained by fitting WLC model to force-extension curves(see supplementary fig 3). This is because polymer is picked at random points along the contour and also the polydisperse nature of polymer. If the curves superimpose well, such a normalization procedure indicate that single molecule event was sampled (see supplementary fig 3).(3) Statistics of such curves obtained was (N$\sim$ 50) for each polymer in their respective solvent conditions. All these curves were fitted to extract model parameters (data not shown). In the paper, representative force-curves (N$\sim$ 5) are shown and normalized for fitting procedure in respective solvents. 
\begin{figure}[h!]
\includegraphics*[width=3.1in]{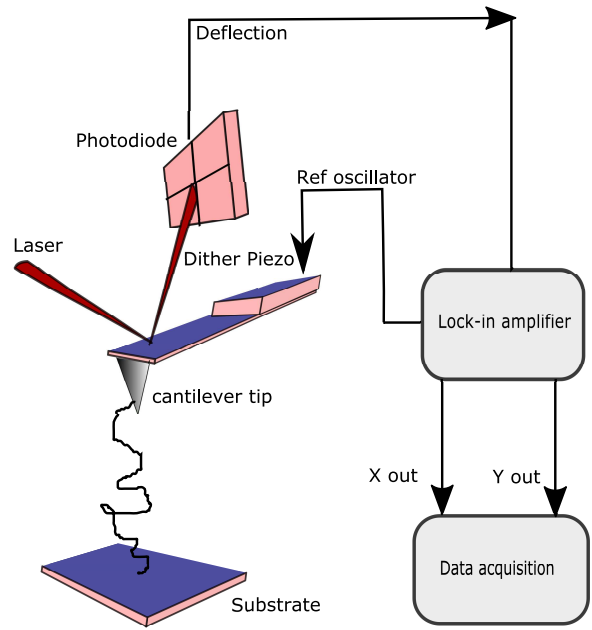}
\caption{ Experimental schematic shows extraction of in-phase (X) and out-of-phase (Y) components of cantilever oscillations using lock-in amplifier with a polymer anchored between the tip and gold substrate. Dither piezo is sinusoidally oscillated with less than 1 KHz frequency and vertical deflection from photodiode is fed into lock-in amplifier. 
}
\label{scheme}
\end{figure}
\subsection*{Dynamic measurement and Analysis}
Dynamic measurements were performed with hyperdrive cantilever holder(Bruker) in which cantilever base is oscillated with a small dither piezo as shown in the schematic of figure 1. A sinusoidal voltage was supplied to dither using the lock-in SR830 (Stanford Research System, Sunnyvale, California) with peak-to-peak amplitude of 1-2 nm. Off-resonance frequencies close to $1$ kHz were exploited (resonance frequency $\sim$ 13 Hz). With sinusoidal oscillation as the reference signal, X and Y component of amplitude from photodiode detector were recorded using lock-in amplifier (Time constant 10 ms). Small amplitudes are necessary for simultaneous comparison with static pulling experiment. Also comparatively low pulling velocity (40-70 nm/s) are needed for narrow bandwidth measurements.
\par
Vibrations of cantilever beam need to be appropriately modelled for correct interpretation of our dynamic measurement.  Flexural vibration of a continuous cantilever beam, can in principle, be decomposed into $n$ eigenmodes. By parameterizing each mode with an equivalent point mass $m^*_n$, stiffness $k^*_n$ and quality factor $Q^*_n$ frequency response (amplitude and phase response) of each mode can be modelled with damped simple harmonic oscillator (SHO). Continuous cantilever beam therefore can be represented as $n$ equivalent simple harmonic oscillators\cite{raman1}.
\begin{figure}[h!]
\includegraphics*[width=2.1in]{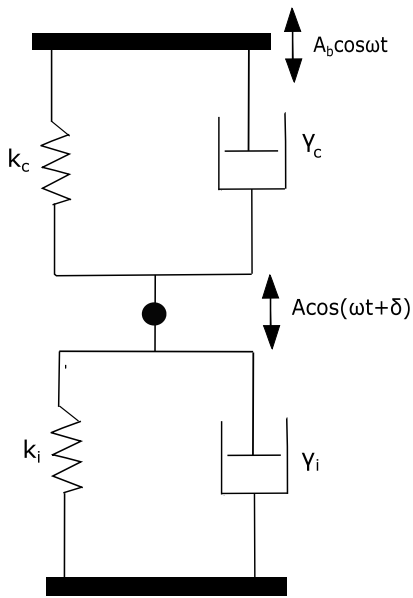}
\caption{Rheological description representing our experimental configuration. It comprises of two voigt-kelvin element(i.e., spring and a dashpot) corresponding to the cantilever and polymer. Two elements are acting in parallel i.e their respective spring and dashpot effectively add with each other. 
}
\label{scheme2}
\end{figure}
In the past, interpretation of dynamic AFM measurement were based on phenomenological point mass model which considers only first eigenmode ($n=1$) to represent the vibrating cantilever. For off-resonance operation employed here, which do not involve eigenmodes, point mass description is appropriate but with a caveat. The oscillating motion of the base is comparable to oscillating amplitude at the cantilever-tip. For quantitative description, base motion need to be explicitly accounted for in the equation of motion because AFM measures deflection or bending with respect to the base.\cite{kulik}. Also, effects of hydrodynamic loading on the lever due to motion of surrounding fluid are included in the equation of motion. $m^*$ is mass that includes an added mass due to inertial loading of fluid moved along with the oscillating lever. $\gamma_c$ is viscous damping of the cantilever in the surrounding fluid. Off-resonance operation allows for small amplitudes of oscillation and therefore linearization of interaction force as $F_{int} = k_{i}y - \gamma_{i}\dot y$ where y is instantaneous cantilever position. $k_i$ and $\gamma_i$ are stiffness and damping of molecule. According to the rheological description\mbox{\cite{pethica,kulik} (Figure \ref{scheme2})}, $k_i$ and $\gamma_i$ are added in parallel with cantilever spring constant $k_c$ and damping $\gamma_c$. This is explained (see also supplementary fig 2) and equation of motion governing cantilever-molecule system that includes the base-motion is
\begin{equation}
 m^*\frac{\partial^{2}y}{\partial t^2}+\gamma \frac{\partial y}{\partial t}+ky = k_cA_b\cos\omega t
\end{equation}

 where $m^*$, $\gamma = \gamma_i + \gamma_c$ and $k = k_i + k_c$ are effective mass, damping and stiffness of molecule plus cantilever system respectively. The term $k_cA_b\cos(\omega t)$ is base motion contribution where $\omega$ and $A_b$ are drive frequency and amplitude of cantilever dithering. Note that because of base motion, the instantaneous cantilever tip-position $y$ is the sum of cantilever deflection with respect to the base $z$ and base displacement $A_b\cos(\omega t)$. Since AFM measures deflection or bending with respect to cantilever base $z$ and not the cantilever position $y$. Hence, transforming equation $1$ with measured $z = y - A_b\cos(\omega t)$ we have,
\begin{equation}
 m^*\frac{\partial^{2}z}{\partial t^2}+\gamma \frac{\partial z}{\partial t}+kz = \underbrace{A_b(m^*\omega^2 - k_i)}_{A\cos\delta}\cos\omega t +  \underbrace{A_b(\gamma\omega)}_{-A\sin\delta}\sin\omega t\\  
\end{equation} 
\begin{equation}
= A\cos\omega t + \delta \notag                                                                                                               
\end{equation}
For off-resonance operation first two terms on left hand side are neglected\cite{sarid}. Assuming $k_i\ll k_c$ and identifying amplitude $z$ as $X\cos\omega t + Y\sin\omega t$, we see from eq (2) that 'in phase' component $X=A\cos\delta$ and 'quadrature' component $Y=A\sin\delta$ of lock-in signal amplitude are directly proportional to elastic and viscous response of molecule respectively. X and Y are: 
\begin{equation}
X={\frac{A_b}{k_c}} (-k_i + m^*\omega^2)
\end{equation} 
and 
\begin{equation}
Y ={\frac{-A_b\omega}{k_c}} { (\gamma_i + \gamma_c  )} 
\end{equation}
%This relation is consistent with elaborate continuum approach that considers solving the standard Euler-Bernoulli equation for vibrations of cantilever beam \mbox{\cite{Deitler2}}. It therefore justifies the use of rheological description in Figure \mbox{\ref{scheme2}} for our experimental configuration. This implies that simple addition of two springs or dashpots have essentially allowed for a simple addition of cantilever and polymer contribution in X and Y.
X and Y in volts are converted to angstrom using dc (static) deflection sensitivity calibrated by the response in deep contact and it is further used for spring constant calibration. Proper quantification of stiffness requires accurate estimate of $A_b$. As expected, for deep contact with hard glass surface lock-in amplitude $\tilde{A}={2\over{3}}L\large{dz\over{dx}}=A_b$ follows the motion of dither piezo and gives an indirect way of measuring $A_b$. However this relation may still involve small fluid contribution because of cantilever base motion\cite{Raman} and therefore for our measurement we employed non-destructive way of directly measuring $A_b$ using home built Fabry- Perot based interferometer detection (see supplementary fig 1). This calibration gives the same value to within 10 \% as given by deep contact. Error in measuring stiffness $k_i$ will primarily arise from systematic error in amplitude $A_b$(supplementary). We believe that a maximum systematic error of 20 \% in $A_b$ will not significantly change our final result. The same procedure as mentioned in force measurement was followed for single molecule detection. The simultaneous dynamic measurements corresponding to such single molecule force events were analyzed. 50 curves for PEG in water and 40 curves for PEG in 2-propanol were analyzed. For polystyrene, 40 and 20 curves in water and 8M urea were analyzed respectively. Representative curves are shown in the paper.

\begin{figure*}[htb!]
\includegraphics*[width=5.5in]{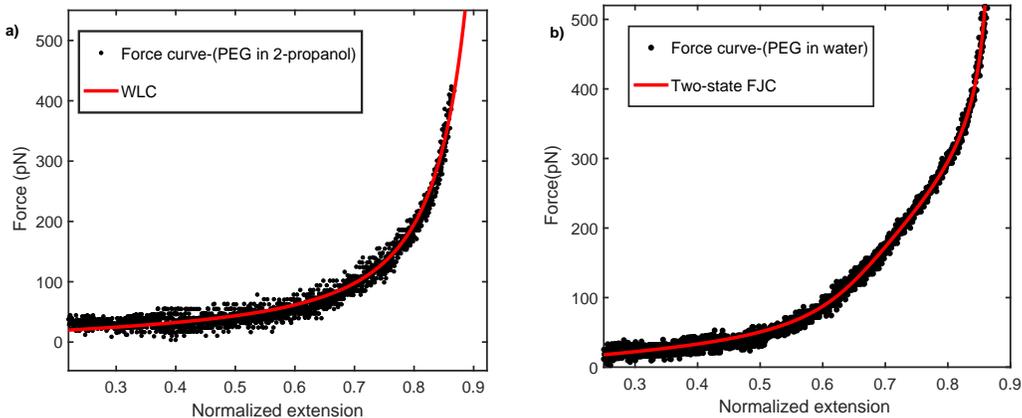}
\caption{a) WLC fit to normalized force-extension curves measured in 2-propanol. WLC fits the force regime with persistence length $l_p=0.13$ nm. b) Two-state FJC fit to normalized force extension curves measured in water. The estimated parameter of kuhn length is $0.24$ nm. Additional best fit parameters include(see supplementay) free energy difference between gauche and trans conformer $\Delta G = 3.1 k_BT$ and length of gauche state $L_{gauche} = 0.23$ nm. }
\label{wlc}
\end{figure*}
\section*{Results and Discussion}
\subsection{Poly(ethylene)glycol}
\subsubsection{Force-extension measurement}
Typical force versus extension curves of polyethylene glycol PEG in 2-propanol recorded in single molecule force spectroscopy experiments are shown in Figure \ref{wlc}a. It shows force versus extension curves normalized by their apparent chain extension(see supplementary fig. 3). At a phenomenological level, these non linear force-extension profiles are described with statistical models like wormlike chain (WLC) or freely jointed chain (FJC). WLC treat the polymer as a continuum chain and accounts for entropy dominated contribution to elasticity with two adjustable fitting parameters, persistence length, $l_p$ and contour length, $L$. Persistence length is the characteristic length-scale beyond which thermal energy can bend the polymer chain. Therefore, it is a measure of local bending stiffness of the polymer. Equation $5$ describes the Marko-Siggia interpolation formula relating force $F$ to extension $x$ of polymer chain which accounts for low F $<$ $k_B$T/$l_p$ and high force F $>$ $k_B$T/$l_p$ behavior of WLC
\begin{equation} 
F={k_BT\over{l_p}}\bigg({1\over{4(1-{x\over{L}})^{2}}}-{1\over{4}}+{x\over{L}}\bigg)
\end{equation}
with boltzmann constant $k_B$ and temperature $T$ in kelvin. Figure \ref{wlc}a shows fitted WLC (eq 5) to normalized force-extension profile. The value of persistence length estimated is $l_p=0.13\pm0.02$ nm. WLC fits reasonably well but persistence length obtained is even lower than the monomer size of $(0.4 $nm$)$. Note that models with additional free parameters (such as extended-WLC) did not change the goodness of fit or the value of persistence length. Measurements were also performed in deionized(DI) water as shown in \ref{wlc}b. PEG in aqueous medium is known to exhibit a conformational change from gauche to trans as a function of force shown in experiments\mbox{\cite{gaub,gaub1}} and simulation\mbox{\cite{heymann}}. The effect of this transition is seen as linear-force region between 100 and 300 pN in \ref{wlc}b, which is not present in 2-propanol. To account for this, force-extension profiles are typically modelled with two-state FJC model\cite{gaub,Cui1}(supplementary eq 1). This model combines FJC entropic elasticity with changes in length of PEG monomers from gauche to trans as force is applied. Figure \ref{wlc}b shows normalized force extension curves fitted with two-state FJC model.
The value of Kuhn length obtained is $b_k = 0.24$ nm which matches well with its value in 2-propanol($2l_p \sim 0.25$ nm) and other organic solvents\mbox{\cite{Cui2}}. However, the Kuhn length $0.24$ nm is about about five time less compared to $1.2$ nm(or persistence length 0.6 nm) measured in low-force magnetic tweezer measurements\mbox{\cite{saleh,saleh2}}. Note that a prior work by Oesterhelt et.al\mbox{\cite{gaub}} chose a value of Kuhn length(0.7-0.8 nm) as measured for PEG in hexadecane, to model PEG elasticity with two-state FJC model in water. However, recent measurement shows that such a large value is likely a result of excluded volume effects due to large size of hexadecane\mbox{\cite{Cui2}}.      

\par
The unphysical value of $l_p$ seem to suggest that WLC model of entropic elasticity is not the adequate description of AFM data, although it fits well since persistence length is a fit parameter. It is suggested that in typical force range of AFM experiments, WLC behavior will transition to FJC-like behavior\mbox{\cite{rosa,rubinstein}}. Based on this, a WLC-FJC interpolation model was proposed which has features of both WLC (persistence length $l_p$) and FJC( a discrete bond length $b$) \mbox{\cite{rubinstein}}. This interpolation  model was used to analyze AFM stretching data on different polymers(see Appendix C,  ref.\mbox{\cite{rubinstein}}). It is observed that the value of persistence length are similar and does not change with respect to original fitting with WLC model.   
%As suggested in ref, the choice between two main class of models i.e  FJC and WLC depends on the range of applied forces and characteristic length scales such as persistence length $l_p$ and effective monomer separation/bond length $b$. At high forces F $>>$ $k_B$T/$l_p$, nonlinear WLC force relation (eq 5) is given by $1-x/L \sim (4Fl_p/k_BT)^{-1/2}$. However, beyond a critical force, intrinsic bending stiffness between consecutive bonds($l_p$) becomes irrelevant and discrete nature of polymer backbone($b$) comes into play. Thus, chain should be viewed as being made up of freely jointed bonds with force relation given by FJC chain $1-x/L \sim (Fb/k_BT)^{-1}$. Comparing the above high force scaling relations gives the WLC to FJC transition force as $F_c \sim k_BTl_p/b^2$.
We tested such a model on our data and similarly no change in persistence length was observed(supplementary fig 4). It suggests that choice of  a  different model  does not explain unreasonably low values of persistence lengths in AFM experiments.
%transition to FJC behavior is not likely in range of forces covered in AFM experiments. Indeed, such a transition is not reported in experiments including low-force($<$ 100 pN) magnetic tweezer measurements\mbox{\cite{saleh3}}. %For synthetic polymers, typical value of persistence length as measured in low-force magnetic tweezer manipulation (or force free condition) is $l_p =0.6$ nm\mbox{\cite{colby,saleh,saleh2}} and C-C bond length $a$ $=$0.15 nm, this transition force is about 500 pN.
This begs the question that why anomaly low values of persistence length( or Kuhn length) repeatedly appear in AFM-based pulling experiments.

Another consideration which possibly explains such anomaly is the method of performing the conventional pulling experiments using AFM. It has been recognized that cantilever probe couples with the polymer elasticity in an intricate manner for constant velocity pulling experiments\mbox{\cite{franco,kreuzer}}. This produces a biased trajectory wherein polymer is not able to sample its intrinsic equilibrium conformations and experiment may not capture the elastic response of the polymer chain alone. This is in contrast to magnetic tweezers experiments wherein properties of isolated polymer in equilibrium are measured\mbox{\cite{kreuzer}}. In the next section, we describe our oscillatory rheology measurements on single polymer chain which deconvolute the effect arising due to cantilever from the polymer's elastic response. We discuss the principle of deconvolution in the last section.

\subsubsection{Dynamic measurement}
\begin{figure}[h!]
\includegraphics*[width=3.2in]{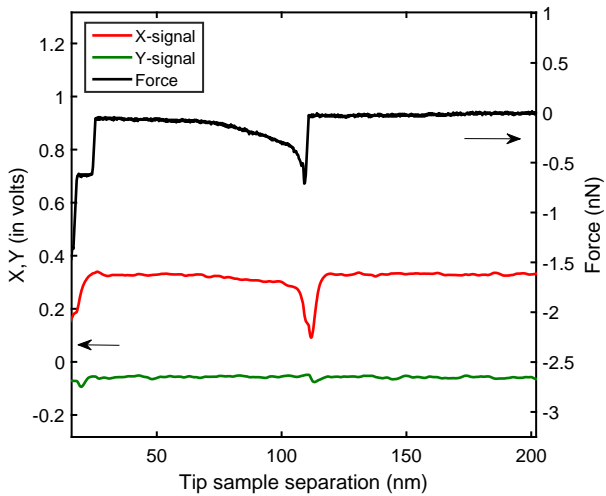}
\caption{ For PEG in water, simultaneously measured raw profiles of force in constant velocity(60 nm/sec) pulling experiment (black) and also in-phase X-signal (red) and quadrature Y signal (green) of lock-in amplifier.
}
\label{raw}
\end{figure}
\begin{figure*}[htb!]
\includegraphics*[width=6.6in]{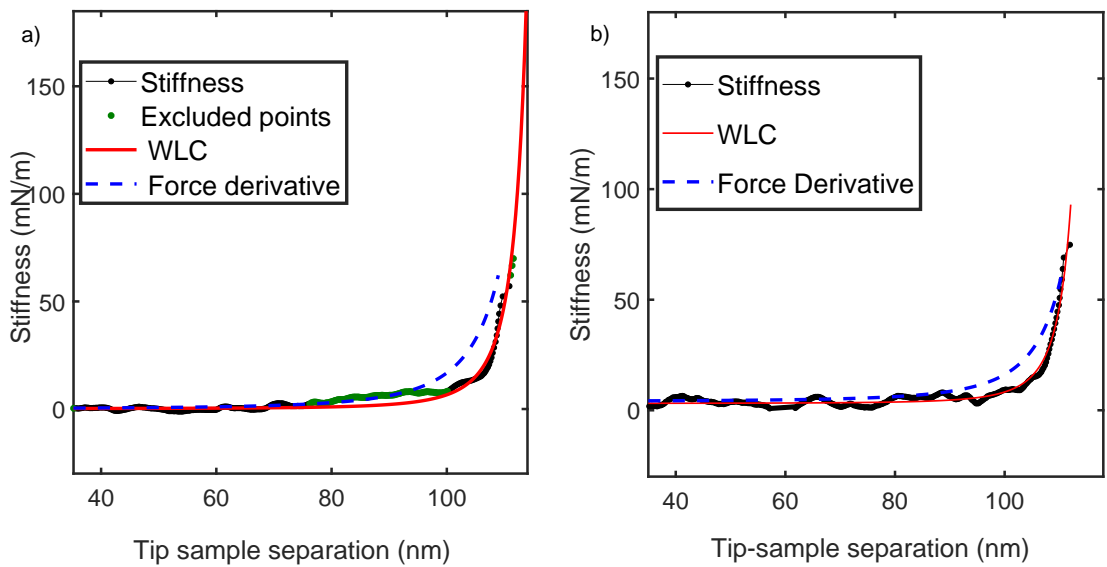}
\caption{Comparison of stiffness obtained via dynamic oscillatory method for PEG with conventional force curve derivative a) in water. Derivative of fitted WLC(blue dash) $l_p=0.13$ nm obtained in constant velocity pulling experiment simultaneous with stiffness from dynamic oscillatory method(black) fitted with WLC derivative(red) $l_p=0.65$ nm. Note that green region between 100 and 300 pN(or extension between 80 and 100 pN) is excluded while fitting to WLC or its derivative. b) in 2-propanol. Derivative of fitted WLC(blue dash) $l_p=0.13$ nm from constant velocity pulling experiment compared with stiffness from dynamic oscillatory(black) method fitted with WLC derivative(red) $l_p=0.64$ nm. 
}
\label{dynamic}
\end{figure*}
To address this, we directly measure elastic response of PEG chain using dynamic oscillatory method along with constant velocity pulling force-extension curve. For this, the cantilever is simultaneously oscillated  while it is pulled slowly at a constant velocity $\sim$ 60 nm/s. This velocity is much less than pulling velocities used in the previous reports\cite{radiom,Cui}.  Using lock-in amplifier we record the dynamic linear response at a frequency close to 1 KHz. Figure \ref{raw} shows raw profile of X and Y signal of lock-in amplifier measured simultaneously with force-extension curve of PEG in water. From eq $3$, in-phase X signal is directly proportional to the elastic response of polymer $k_i$ apart from a fixed contribution from cantilever's hydrodynamic loading. X-signal follows the feature of force-extension curve and subtracting the fixed cantilever contribution one extracts the elastic response of polymer. Quadrature Y signal is proportional to effective friction of the combined polymer cantilever system. Y signal shows no difference with and without polymer attached and remains featureless in Figure \ref{raw}. In our recent work, we have shown that the dissipation due to unfolded protein chain of I27 is also immeasurably low\cite{shatruhan}. This is possibly due to much faster polymer internal dynamics compared to experimental timescales\cite{moglich}. In Figure \ref{dynamic}, the stiffness of the polymer extracted using the in-phase X component data is shown along with derivative of WLC fit to force-extension data. We chose the derivative of the fit for comparison since the derivative of experimental data become noisy for any meaningful comparison. The derivative of force-extension behavior(blue dash) shows marked deviation from directly measured stiffness data(black)for PEG in water(Figure \ref{dynamic}a). Analyzing the stiffness data with WLC model(red) gives persistence length of $l_p=0.65\pm 0.19$ nm compared to $l_p=0.13\pm 0.02$ nm from derivative of force-extension curve. We excluded part of the data in the extension range of 80 to 100 nm (green) which show marked deviations while fitting both force and stiffness data to WLC (see supplementary fig 5). As explained earlier, PEG monomers undergo a well-known trans-trans-gauche to all-trans conformational change in water as shown in experiments\mbox{\cite{gaub,gaub1}} and simulation\mbox{\cite{heymann}}. As observed\mbox{\cite{gaub1}}, this transition characteristically shows up as a linear force region between 100 and 300 pN(80 and 100 nm). Deviation of stiffness data in similar force region(80 and 100 nm) shows that our dynamic measurement is sensitive to such conformational change. However, due to conformational change such a fitting procedure with WLC cannot be used to extract any meaningful estimate of persistence length.
To avoid the conformational change in water, we performed measurements in 2-propanol which is also a good solvent for PEG. Since 2-propanol is volatile, the experiment is performed within 2 hour of injecting it into fluid cell. The effects due to conformational change are now absent in both the dynamic stiffness in Figure \ref{dynamic}b) as well as in constant velocity pulling data (figure \ref{wlc}a). For PEG in 2-propanol, direct measurement of stiffness through oscillatory rheology gives the persistence length $l_p=0.64\pm 0.19$ nm and WLC fits well in the entire range without having to exclude any region due to hydration effects as in Figure \ref{dynamic}a. The constant velocity pulling experiments yield a value persistence length($0.13\pm 0.02$ nm).   
Persistence length of $0.64\pm 0.19$ nm is consistent with low-force equilibrium measurements performed using magnetic tweezers manipulation\cite{saleh,saleh2} and also other ensemble techniques\cite{colby,colby2}. Dittmore et. al\cite{saleh} and Innes-gold et. al\cite{saleh2} reported the $l_p$ value of $0.55$ nm and $0.65$ nm respectively. And the neutron scattering measurements by Smith et. al\cite{colby2} report the $l_p$ for PEG as $0.6$ nm. 
\par
In the context of extracting viscoelastic information, oscillatory measurements have been performed on single molecules\mbox{\cite{kawakami,kawakami2,liang}}. In contrast to the present study, these measurements were performed with oscillation frequencies close to cantilever resonance and have reported dissipation for single molecule. Recent work in modelling of cantilever dynamics in liquid, highlights the problem in interpretation of single-molecule dissipation data\mbox{\cite{Deitler2,shatruhan}}. It points out difficulty in making a distinction between elastic and dissipative signals especially for frequencies close to cantilever resonance.
Dynamic measurement on PEG  were also performed in water previously\mbox{\cite{hinterdorfer}}. Kienberger et al.used PEG derivatives with functional groups at the chain ends to couple the chain to tip and the substrate and  magnetic excitation scheme to oscillate the cantilever close to resonance. The measured static pulling as well as dynamic force-extension profiles were analyzed with the extended Worm-Like-Chain (WLC) model, yielding a persistence length of 0.38 nm. These measurements are not strictly off-resonance as compared to ours. The use of magnetic excitation method, however are known to produce artefact-free measurements\cite{raman2}. It is noteworthy that the oscillatory measurements in their case also  produce reasonably good persistence length estimates. A detailed investigation to  compare  our method to the one used by Kienberger et al. is  worthwhile. 

In the next section we describe our measurements on polystyrene(PS) in good and poor solvents. This sheds light on the influence of solvents on our dynamic measurement and ultimately understanding its effect on polymer elasticity.
\subsection{Polystyrene}
In order to further examine the effect of poor solvents on  dynamic oscillatory measurements, we performed force spectroscopy measurement on polystyrene in water. One difference between the two homopolymers is that PS has aromatic side chains which are hydrophobic and the polymer chain collapses in poor solvent such as water. It has been demonstrated recently that high concentration aqueous solution of urea acts as good solvent for PS \cite{zangi,mondal,england}. We performed both,  the constant velocity pulling and dynamic oscillatory measurements, on PS in water and PS in 8 M urea solution in water. 

\subsubsection{Force-extension measurement}
\begin{figure}[h!]
\includegraphics*[width=3in]{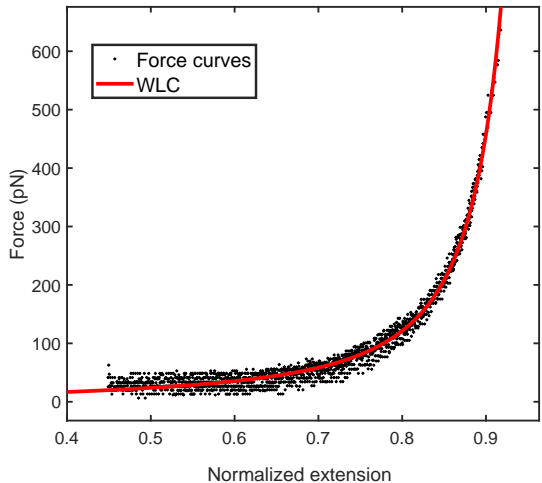}
\caption{Normalized force-extension curves for polystyrene in water with persistence length $l_p=0.23$ nm.
}
\label{polyst_wlc}
\end{figure}
\begin{figure*}[bt!]
\includegraphics*[width=6in]{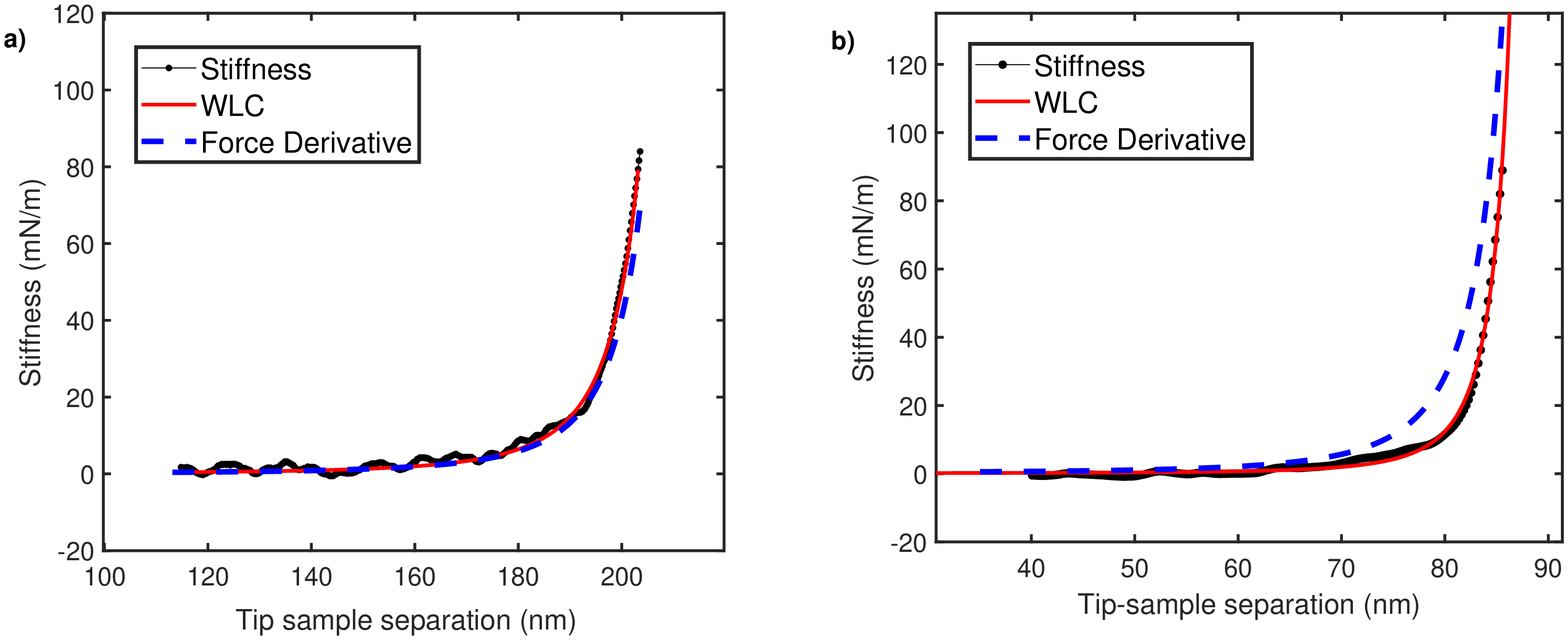}
\caption{Comparison of derivative of WLC obtained from fitting to force-extension curve(blue dash), stiffness measured with dynamic oscillatory(black) method and WLC derivative fit(red) to stiffness. a)Polystyrene in water b)Polystyrene in 8M urea. 
}
\label{polyst}
\end{figure*}
%Water is an important physiological medium for proteins and other biomolecules. For instance, an unfolded polypeptide due to hydrophobic side chains collapse in water to form a globule. This step is thought to be critical in driving self-assembly of globular proteins in protein folding. Polystyrene, due of its homogeneous and simple chemical structure, is used as a model hydrophobic polymer to study collapse in gobular proteins and in general hydrophobic hydration in polymers\cite{li2}.

Figure \ref{polyst_wlc} shows normalized force extension profiles fitted with WLC. Note that we fit in the entire force regime ignoring the flat plateau region ($<$20 pN) which results from coil-globule transition for polystyrene in water\cite{walker2}. The fit gives a persistence length $l_p=0.23\pm0.02$ nm consistent with observation made earlier\cite{walker}. The steric hindrances are offered by aromatic side group of benzene. It is 0.72 nm long and is responsible for determining the persistence length. In this sense, the size of the monomer is about $0.72$ nm and the  observed  low value of persistence length has been debated\cite{radiom}. It is expected to have a $l_p$ larger than $1$ nm. Low value of $0.25$ nm is also consistently cited in ideal to good solvents like toluene\cite{radiom,walker2}.
\subsubsection{Dynamic measurement}
We repeat our simultaneous oscillatory measurement along with force-extension measurements on polystyrene in water. Figure \ref{polyst}a shows stiffness measured by dynamic oscillatory method(black), WLC fit to it (red) and derivative of WLC fit to the force-extension curve (blue dash). Interestingly, they  coincide with each other having no clear deviation as in case of PEG in water or 2-propanol. Persistence length obtained by fit to the dynamic stiffness data is similar $l_p=0.21\pm 0.02$ nm. Similarly, in contrast to our observation on PEG in good solvents(\ref{dynamic}), the unfolded globular proteins show same value of persistence length in AFM\cite{Lorna}, MTs\cite{popa} and also oscillatory measurements\cite{Deitler2,shatruhan}. As in case of polystyrene, we suspect that fitting WLC to the data yields persistence length as only an effective parameter. However, in WLC model the strong hydrophobic effects are not considered along with backbone entropic contributions. A need to include this effect has already been suggested in the literature\cite{grater,berne}.
\par
In order to confirm this effect, we employed our measurement scheme on polystyrene in good solvent. High concentration of aqueous urea solution  behaves similar to a good solvent for hydrophobic polymer, as it weakens the hydrophobic interaction which causes the polymer to collapse in water\cite{zangi,mondal,england}. We carry out dynamic measurement in 8M urea as shown in Figure \mbox{\ref{polyst}b}. It shows the comparison of dynamic and constant velocity pulling measurements. Interestingly, we obtain a similar deviation between the derivative of WLC fit to force-extension curves($l_p$=0.25 nm) and stiffness-extension curves measured using dynamic oscillatory method ($l_p$ as 0.8 nm). This value of $l_p$ is reasonable since it is larger than monomer size. The similarity in results between PEG  and PS in good solvent is striking. We discuss it further in the next section. We argue that the  hydrophobic interactions dominate the polymer elasticity in such a way that stiffness from dynamic oscillatory technique and derivative of force extension curves approach each other.

\subsection{Principle of deconvolution}
To qualitatively explain the deviation between derivative of force-extension curve and oscillatory measurement we begin by noting that conventional constant velocity pulling experiments represent a coupled cantilever-molecule system. In AFM experiments, the extension of molecule and force on it are measured indirectly through the deflection of cantilever probe to which the molecule is tethered. This deflection force is determined by simultaneous action of both cantilever and the molecule in an intricate convolution and effect of one from the other is difficult to separate\mbox{\cite{franco,kreuzer}}. To understand this better, one makes a connection between force-extension curve and polymer stretching thermodynamics(see supplementary information). Under equilibrium condition, canonical partition function $Z_{system}$ of composite cantilever plus molecule system at a given displacement $D$ of cantilever at time t can be written as\mbox{\cite{franco,kreuzer}}:
\begin{equation}
    Z_{system}(D) \sim Z_m*Z_c= e^{{-\beta F(x)}} e^{{-{\beta k_c\delta^2}/2}}
\end{equation}
where $\beta$ is 1/$k_BT$, $x$ is end-to-end extension of molecule. $Z_m$ is the partition function of isolated molecule determined by its molecular free energy F($x$) and $Z_c$ is cantilever partition function determined by its stiffness $k_c$ and deflection $\delta$. Because deflection is $\delta$=$D-x$, the function in eq (6) defines the mathematical convolution of cantilever and molecule in pulling experiments. Therefore,  measured quantities such as average force determined by $Z_{system}$ are those of composite molecule plus cantilever system(see supplementary). However, one is interested in the properties of molecule itself determined by $Z_m$. One notes from eq(6) that either at high force characterized by large deflection or the use of stiff spring $k_c$, contribution of cantilever biasing potential $k_c(D-x)^2/2$ is reduced substantially with $Z_c$ approaching delta function $\delta(D-x)$\mbox{\cite{kreuzer}}. This implies that at high forces the derivative of force-extension curve and stiffness from the dynamic measurement will start to coincide as seen in fig 5a, 5b and fig 7b. Similarly, for soft spring or low force magnetic tweezer measurement done at a constant force,  partition function reduces to that for an isolated molecule\mbox{\cite{kreuzer}}. Therefore, effect of coupling would be maximum at intermediate forces. To see this, it is noted that a free energy($F$) in eq(6) is determined by WLC entropy ($-TS_{WLC}$). At an intermediate force ($\sim$ 250 pN), cantilever potential is $\sim$ 12 kT compared with WLC entropic contribution($-TS_{WLC}$) of about 2 kT( see supplementary information). Therefore, from eq(6), cantilever dominantly determines the overall $Z_{system}$, producing a biased force-extension curve. In contrast, directly measuring stiffness by externally oscillating the cantilever, stretched polymer and the cantilever work in parallel pathways. This means that the contribution of polymer and cantilever to the overall system response is simply additive as seen from eq(3) and eq(4). Therefore, oscillatory measurement allows to extract polymer intrinsic response from its coupling to cantilever probe.
For externally modulated AFM cantilever it is known that spring for the polymer effectively adds with cantilever to determine net force on the system\mbox{\cite{pethica,kulik}}(supplementary fig 2). This observation results in parallel coupling pathway in eq (3).

In addition to conformational entropy of backbone, polymer-solvent interactions significantly influence elasticity of extended hydrophobic polymers like polystyrene\mbox{\cite{grater,berne2,radiom,gosline}}. It is well established that hydrophobic interactions among nonpolar monomers tend to soften the polymer, reducing its persistence length by a large factor\mbox{\cite{grater,berne2}}. Thermodynamically, the strength of hydrophobic interactions can be described in terms of hydrophobic solvation free energy between polymer and solvent. Exposure of hydrophobic side chains upon extension would require hydrophobic free energy penalty that is large and positive\mbox{\cite{gosline}}. It is estimated from experiments to be about six times the chain entropy\mbox{\cite{grater}} even for relatively large chain extension and lowers the persistence length significantly. With bulky aromatic side chain, hydrophobic free energy per monomer of polystyrene is estimated to be $\sim$ 20 kT\mbox{\cite{abraham}}. Hence, from eq (6) we see that total free energy F($F_{hydrophobic}$+$F_{WLC}$) starts to become dominant compared to cantilever biasing in determining overall $Z_{system}$. Therefore, the dynamic oscillatory method and the constant velocity pulling experiments produces similar stiffness profiles for polystyrene in water. Since WLC only accounts for entropic contributions, one should avoid its use to analyse the data of both unfolded proteins and hydrophobic polymers in poor solvents.

\section*{Conclusions}
We employ oscillatory rheology to get an independent yet simultaneous estimate of elastic response in single molecule pulling experiments. This response shows significant deviation in elasticity from conventional constant velocity pulling experiments in good solvents. The  analysis with WLC model, which accounts for entropic nature of elasticity, results in a large and physically acceptable value of persistence length. This value is consistent with equilibrium magnetic tweezers measurements, which suggests that oscillatory dynamic measurement using AFM is better suited for measurement of single polymer chain elasticity compared to 
constant velocity pulling experiment. We conclude that constant velocity pulling experiments using AFM produces a biased molecular trajectory due to convolution of cantilever probe with molecule's response.  
In contrast, for polymer chain in poor solvent no deviation is observed in stiffness profiles extracted from data using both methods. We explain this by considering the contribution of additional hydrophobic free energy and its effect on sampling the intrinsic equilibrium configuration in poor solvents. Such considerations explain our observation that  both methods yield the same elastic response for polymer single chain in poor solvent.   

While pulling experiments were suspected to be biased in earlier works, present study is the first to show oscillatory technique to deconvolute apparatus effects. Although we use WLC at a phenomenological level, a persistence length consistent with other techniques underlines its success in high-force AFM stretching experiments. However, results hint at observing caution for its use in describing AFM data of polymers in poor solvents.
 
\section*{Supporting Information}
\begin{itemize}
    \item Calibration of drive amplitude of dither piezo; Error analysis; Normalization of PEG force-extension curves; Stiffness add in parallel; two-state fjc, wlc-fjc interpolation; Connection between force-extension curve and thermodynamics.
    \end{itemize}

\section*{Acknowledgements}
The work was supported by Wellcome Trust-DBT India Alliance fellowship to SP (500172/Z/09/Z). Amrita Kulkarni helped in preparing samples.  VA and SR would like to acknowledge fellowship from IISER Pune.

\end{document}